\documentclass{rsproca}

\usepackage{graphicx,subfigure,amsmath,bbm}
\usepackage{tabularx}
\usepackage[square,comma,sort&compress,numbers]{natbib}

\jname{rsif}
\Journal{J. Roy. Soc. Interface }

\begin{document}

\title{Nonclassical phase diagram for virus bacterial co-evolution mediated by CRISPR}

\author{
Pu Han$^{1}$ and Michael W. Deem$^{1,2,3}$}

\address{$^{1}$Department of Physics \& Astronomy, Rice University,
Houston, TX 77005, USA\\
$^{2}$Department of Bioengineering, Rice University, Houston, TX 77005, USA\\
$^{3}$Center for Theoretical Biological Physics, Rice University, Houston, TX 77005, USA}


\keywords{CRISPR, phase diagram, bacteria, virus, extinction, co-evolution}

\corres{Michael W. Deem\\
\email{mwdeem@rice.edu}}

\begin{abstract}
CRISPR is a newly discovered prokaryotic immune system. Bacteria and archaea with this system incorporate genetic material from invading viruses into their genomes, providing protection against future infection by  similar viruses.
The conditions for coexistence of prokaryots and viruses is an interesting problem in evolutionary biology.
In this work, we show an intriguing phase diagram of the virus extinction probability, which
is more complex than that of the classical predator-prey model. As the CRISPR incorporates genetic material, viruses are under pressure to evolve to escape the recognition by CRISPR. When bacteria have a small rate of deleting spacers, a new parameter region in which bacteria and viruses can coexist arises, and it leads to a more complex coexistence patten for bacteria and viruses.
For example, when the virus mutation rate is low, the virus extinction probability changes non-montonically with the
bacterial exposure rate.
The virus and bacteria co-evolution not only alters the virus extinction probability, but also changes the bacterial population structure. Additionally, we show that recombination is a successful strategy for viruses to escape from CRISPR recognition when viruses have multiple proto-spacers, providing support
for a recombination-mediated escape mechanism suggested experimentally.
Finally, we suggest that the reentrant phase diagram, in which phages can
progress through three phases of extinction and two phases of abundance at
low spacer deletion rates as a function of exposure rate to bacteria, 
is an experimentally testable phenomenon.

\end{abstract}

\maketitle

\section{Introduction}

Clustered Regularly Interspaced Short Palindromic Repeats (CRISPR) is a recently discovered immune system of prokaryotes. It is widely distributed in bacteria and archaea. Nearly half of bacteria and almost all archaea possess the CRISPR system \cite{Grissa2007, Sorek2008, Oost2009, HovathBarrangou2010}. CRISPR is adaptive and heritable 
\cite{Barrangou2007}: bacteria can acquire a short piece of invading DNA (termed proto-spacer) and integrate this piece of exogenous DNA into the CRISPR locus. The nucleotide sequence in the CRISPR locus that  originated from the invading DNA is called a spacer.
The mechanism of the CRISPR system is categorized into three stages: the acquisition and integration of new spacers into CRISPR, expression and maturation of CRISPR RNAs (crRNAs), and CRISPR interference \cite{Marraffininat2010,Makarova2011,Wiedenheft2012}. In the acquisition stage, proto-spacers from viruses (phages) or plasmids are integrated into the CRISPR locus. During the expression stage, CRISPRs are first transcribed to precursor CRISPR RNAs (pre-crRNAs). Pre-crRNAs are then catalyzed by Cas (CRISPR-associated) proteins into mature crRNAs. In the interference stage, crRNAs guide Cas proteins to cleave the complementary DNA of invading plasmids or phages \cite{Monica2009,Swarts2012,Deveau2010,Przybilski2011,Lintner2011,Zhang2012}.

The discovery of the CRISPR system challenged our understanding of
the evolutionary dynamics
of bacteria and phages \cite{Barrangou2007,Banfield2009}.
Several models were established to explain the interesting features of CRISPR
and the co-evolution of prokaryotes and phages. Levin used an ecological model to investigate the question of why and how CRISPR is established and maintained in a bacterial population \cite{Levin2010}. A similar model that considered the conjugational transfer of beneficial plasmids suggested that plasmids may be more likely to evade CRISPR-Cas immunity by inactivation of functional CRISPR-Cas rather than by mutation of the target proto-spacers \cite{Jiang2013}.
He and Deem introduced a population dynamics model to explain the heterogeneous distribution of the spacer diversity in CRISPR \cite{DeemPRL2010}, i.e.\ the decrease of spacer diversity with distance from leader. A later paper considered a density-dependent phage growth model and showed that recombination allows viruses to evade CRISPR more effectively than does point mutation alone when greater than one mismatch between the crRNA and protospacer was required for viruses to escape CRISPR recognition \cite{Pu2013}. Childs et al.\ used an eco-evolutionary model of CRISPR with imperfect immunity to also show that both bacteria and phages were highly diversified by co-evolution and that diversity decreased with distance from leader \cite{Childs2011}. In another paper, a metric, population-wide distributed immunity (PDI), was defined to quantify the immunity distributed among the host-viral population. This model showed that the number of viral proto-spacers, mutation rate, host spacer acquisition rate, and spacer number could change the host-viral population structure by a distributed immunity \cite{Childs2014}. Haerter et al.\ considered spatial effects. Their model showed that CRISPR and spatial self-organization stabilized the coexistence of bacteria and phages. Protected by CRISPR, bacteria could coexist with phages even when the diversity of phages was large \cite{HaerterSneppen2011}. In a follow-up paper, the fitness cost of spacers was taken into consideration. Due to the spatial inhomogeneity and the fitness cost of spacers, it was observed that evolution favors an intermediate number of spacers \cite{HaerterSneppen2012}. Weinberger et al.\ combined a population-genetic model with metagenomic sequencing to study the population dynamics of bacteria and phages \cite{WeinbergerPLosCB2012}. They reported the gradual loss of bacterial diversity through selective sweeps in the host population. This model also showed that the trailer-end of the spacer array was conserved even though the old spacers did not provide immunity against current phages. Increasing the spacer deletion rate repressed the bacterial immunity and led to a viral bloom. Weinberger et al.\ also examined why CRISPRs are more common in archeae than in bacteria \cite{Weinberger2012} with stochastic model of viral-CRISPR co-evolution. The model showed that a decreased viral mutation rate increases the prevalence of CRISPR in archae, and CRISPR appeared only at an intermediate level of innate immunity. In a follow-up paper \cite{Iranzo2013}, a model with explicit population dynamics showed that CRISPR was ineffective for extremely large populations. Because mesophiles usually have larger population sizes, this model gave another explanation for the increased prevalence of CRISPR in hyperthermophiles compared to mesophiles. Finally, a phase diagram of bacteria and phages has been computed, with results similar to the classical predator-prey model \cite{Iranzo2013}, i.e.\ bacteria and phages coexist only when the virulence of phages is not too high and the immunity of bacteria is not too strong.  The mean-field assumption of this approach, however, is in contrast to the strong stochastic effects seen in experiments \cite{Paez2013}.


Here we investigated the conditions under which bacteria and phages can coexist in a fully stochastic model of co-evolution. The competition here differs from that in the classical competitive exclusion principle \cite{Gause_1934,Garrett_Science}, which studies the competition between species that occupy the same ecological niche. In our model, bacteria and phages do not occupy exactly the same ecological niche but rather can co-exist. Phages can hijack bacteria, and bacteria can gain immunity to avoid being infected. 
We studied the impact of different phage evolution strategies, namely point mutation and recombination, on the co-evolution of bacteria and phages.
We found an interesting phase diagram of the extinction probability of phages, which cannot be explained by the classical predator-prey model. In the classical predator-prey model, bacteria and phages only coexist within one parameter region. Outside this region, bacteria and phages cannot both coexist. In this paper, we find bacteria and phages can coexist in several parameter regions.
Indeed, bacteria and phages coexistence is re-entrant as a function of the exposure rate of phages to bacteria, for low phage mutation rates.

\section{Method}

We used a stochastic model to study the population dynamics of bacteria and phages. 
The bacteria have a rate of acquiring and losing spacers. The phages have multiple proto-spacers that can evolve by point mutation and recombination. Spacers and proto-spacers are expressed as a bit string.
Each bit can be either ``0'' or ``1''. The length of each spacer and proto-spacer is $L$ bits. The number of proto-spacers in phages is $n_p$. CRISPR suppresses the phages, and unrecognized phages can infect and reproduce in bacteria. The co-evolving dynamics is described by seven events:

\begin{enumerate}
    \item 	\textbf{Bacteria reproduction}: The growth rate of wild type bacteria that do not acquire any spacers is $c_0$. Each spacer has a cost $c$. The growth rate of bacteria that have spacer array $\vec{s}$ is
        $[1-(x^B+x^I)/x^B_{_\mathrm{M}}]\cdot c_0/(1+c\cdot k_{\vec{s}})$, where $k_{\vec{s}}$ is the number of spacers in the spacer array $\vec{s}$, $x^B$ is the density of healthy bacteria, $x^I$ is the density of infected bacteria, and $x^B_{_\mathrm{M}}$ is the maximum density of bacteria.
    \item   \textbf{Bacteria infection}: Healthy bacteria can be infected by phages. The adsorption rate of phages to healthy bacteria is $\beta x^P  x^B$, where $\beta$ is the exposure rate, $x^P$ is the density of free phages, and $x^B$ is the density of healthy bacteria. Bacteria have a probability $\gamma$ to acquire a new spacer from the invading phage genome. Each proto-spacer has probability $\gamma/n_p$ to be acquired. The newly acquired spacer is always inserted at the leader-proximal end of CRISPR, and the phage is degraded. Old spacers are shifted to the distal end. The maximum number of spacers per bacteria is $n_s$. If the number of spacers reaches $n_s$, the oldest spacer is deleted when a new spacer is acquired. The alternative event, with probability $1-\gamma$, is no incorporation of a proto-spacer. In this case, if any spacer in the CRISPR matches any proto-spacer of the phage, the phage is killed. Otherwise, this bacterium becomes infected.
    \item \textbf{CRISPR deletes one spacer}: A bacterium that possesses the spacer array $\vec{s}$ has a rate
        $d\cdot k_{\vec{s}}$ to delete one spacer, where $k_{\vec{s}}$ is the number of spacers in spacer array
        $\vec{s}$, and $d$ is the rate of deleting one spacer. When one spacer is deleted, the other spacers will be shifted towards the leader end.
    \item	\textbf{Bacterial lysis}: Infected bacteria have a rate $1/\tau$ to lyse, where $\tau$ is the latent time. When the infected bacteria lyse, $b$ new phages come out, where $b$ is the burst size. Each of the newborn phages can have point mutations or recombination.
    \item \textbf{Phage mutation}: Phages upon bacterial lysis can have point mutation. The rate of point mutation is $\mu$ per base per replication. A mutation flips the value of a nucleotide.
    \item \textbf{Phage recombination}: Phages upon bacterial lysis can have recombination. The rate of recombination is $\nu$. A recombination occurs with another phage randomly in the whole phage population, as a mean-field approximation to multiple infection. The recombination crossover probability is $p_c$\cite{Pu2013}.
    \item 	\textbf{Phage degradation}: Each phage has a decay rate $\delta$.
\end{enumerate}

Initially, no bacteria have spacers. There are one or more strains of phages in the environment initially. Each strain of phages has $n_p$ distinct proto-spacers.

The values of parameters are determined by the experimental data (see Supplementary Information). We used the Lebowitz-Gillespie algorithm \cite{Bortz1975,Gillespie1977} to sample the stochastic process of the co-evolution of bacteria and phages. The master equation of this stochastic process is in the Appendix.

\section{Results}

\begin{figure}[htb!]
\includegraphics[scale=0.25]{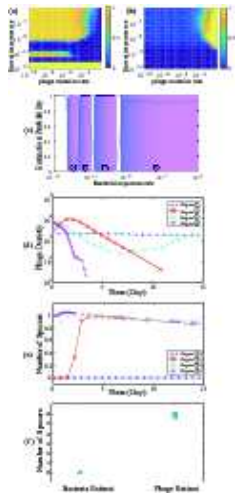}
\caption{\label{extinction1} (Color Online) The extinction probability of bacteria and phages when $\gamma = 0.0005$ and $d = 10^{-5}\ \mathrm{min}^{-1}$. The values of other parameters are $c_0 = 0.005\ \mathrm{min}^{-1}$, $c = 0.1$,
$b=100$, $\tau=40\ \mathrm{min}$, $\delta = 0.001\ \mathrm{min}^{-1}$, $\nu = 0$, $n_p = 30$, $L=10$, $n_s=6$, and  $x^B_{_{\mathrm{M}}}= 10^7\ \mathrm{mL}^{-1}$. The volume of the system in our simulation is $V=10^{-3}\ \mathrm{mL}$. There is one strain of phages initially. The initial bacterial density is $5 \times 10^6\ \mathrm{mL}^{-1}$. The initial phage density is $5 \times 10^7\ \mathrm{mL}^{-1}$.
\textbf {(a)} The extinction probability of phages.
\textbf {(b)} The extinction probability of bacteria.
\textbf {(c)} The extinction probability of phages when $\mu=10^{-8}$ per base per replication in Fig.~\ref{extinction1}(a).
\textbf {(d)} The typical behavior of a density of phages in the first four regions of Fig.~\ref{extinction1}(c).
\textbf {(e)} The average number of spacers in bacteria in the first four regions of Fig.~\ref{extinction1} (c). \textbf{(f)} The average number of spacers in bacteria when bacteria go extinct and when phage go extinct.}
\end{figure}

We examine the coexistence of phages and bacteria as a function
of the phage mutation rate and bacterial exposure rate.
Fig.~\ref{extinction1} shows a phase diagram for the phage and bacterial populations.
In Fig.~\ref{extinction1}(c), there are four transitions in the extinction probability of phages when the mutation rate of phages is small. In region (1), phages begin to emerge but the density of phages stabilizes at a low level. Bacteria and phages can coexist in this region. In region (2), the density of phages increases initially but then decreases to zero. In this region, phages have a high probability to go extinct. In region (3), the density of phages initially increases  and then decreases, but in contrast to the behavior in region (2), phages can grow back and avoid extinction in this case. In region (4), phages rapidly go extinct after a sharp initial burst. The extinction probability of phages is high, and the extinction probability approaches a limit. In this region, bacteria and phages cannot coexist.

The four transitions for the extinction probability of phages as a function of $\beta$ can be explained by Eq.~\ref{case1} and Eq.~\ref{case3}. In region (1), because the density of phages is low and the value of $\beta$ is small, the number of spacers in bacteria is almost 0, Fig.~\ref{extinction1}(e). Therefore, almost all bacteria are susceptible to phages. The equations of infected bacteria and phages can be approximated as
\begin{subequations}\label{case1}
\begin{eqnarray}
\frac{d x^I}{dt} & = & \beta x^P x^B - \frac{x^I}{\tau} \\
\frac{d x^P}{dt} & = & \frac{b}{\tau} x^I - \beta x^P x^B - \delta x^P
\end{eqnarray}
\end{subequations}
where $x^I$ is the density of infected bacteria and $x^P$ is the density of phages. Solving Eq.~\ref{case1}, we find when $\beta^* = \delta/[x^B(b-1)] \approx 10^{-12}\ \mathrm{mL} \cdot \mathrm{min}^{-1}$ the replication rate of phages begins to exceed the phage decay rate, so phages emerge in the system.

In region (2), as $\beta$ increases, the density of phages rapidly increases and bacteria begin to acquire spacers, Fig.~\ref{extinction1}(d) and (e). We can estimate the selection pressure on bacteria in this region. When $x^P \approx 10^9 \ \mathrm{mL}^{-1}$, which is the typical density of phages before bacteria acquire spacers in region (2), the infection rate of each bacterium that has no spacers is $\beta x^P \approx 10^{-3}\ \mathrm{min}^{-1}$, which is the same order as the growth rate of bacteria. So the bacteria that acquire spacers dominate the bacterial population in a short time, and the density of phages will go down, eventually to zero.

In region (3), the phages increase first, then bacteria acquire spacers, leading the phages to decrease, which is similar to the behavior in the region (2). But when the density of phages is low, bacteria will delete spacers due to the deletion rate and the cost of spacers. Because the mutation rate of phages is small, bacteria that acquire one or more spacers have immunity against most phages. Phages can only infect those bacteria that lost all spacers. Here we define the proportion of bacteria that have lost all spacers as $q$. Then the density of susceptible bacteria is $x^B\cdot q$. Thus the equation of infected bacteria can be approximated as
\begin{equation}\label{case3}
\frac{d x^I}{dt}  =  \beta x^P x^B q - \frac{x^I}{\tau}
\end{equation}
In region (3), the value of $q$ is roughly $0.1$ from Fig.~\ref{extinction1}(e), so we can find $\beta^* = \delta/[x^B(q b-1)] \approx 10^{-11}\ \mathrm{mL}\cdot \mathrm{min}^{-1}$. Therefore, in region (3), phages can grow back when some portion of bacteria lose spacers. As the density of phages increases, the average number of spacers in bacteria also increases, which in turn represses the growth of phages, as in Figs.~\ref{extinction1}(d) and (e). So in this case, the density of phages fluctuates around a low value and eventually stabilizes.

The density of free phages decreases due to two factors. One factor is decay. The other factor is due to CRISPR recognition and subsequent degradation. Therefore, the overall decay rate of phages is $\beta x^B + \delta$. In the left boundary of region (4), $\beta$ is the order of $10^{-9}\ \mathrm{mL}\cdot \mathrm{min}^{-1}$, and the overall decay rate of phages is $\beta x^B + \delta \approx 10^{-2}\ \mathrm{min}^{-1}$. Following the same argument as in region (3), the minimum value of $q$ for which phages can grow back is $q^* = (\beta x^B + \delta)/(b \beta x^B) \approx 10^{-2}$.
The time required for $q^*$ bacteria to lose spacers is $t > q^* / d \cong 1000$ min, which is longer than the half life of phages. Thus, before bacteria lose spacers, all of phages are adsorbed into bacteria. Because bacteria have acquired spacers and the mutation rate of phages is small, phages that are adsorbed into bacteria are killed by CRISPR. Therefore, in region (4), phages go extinct rapidly after the initial burst. When $\beta$ further increases, if bacteria acquire spacers, phages will go extinct rapidly. If bacteria do not acquire spacers, bacteria will go extinct, as in Fig.~\ref{extinction1}(f). The extinction probability of phage approaches a limit, $1-(1-\gamma)^{N^B_0}\approx 0.918$ in Fig.~\ref{extinction1}, the probability that one of the initial bacteria acquired a spacer, where $N^B_0$ is the initial bacterial population.

\begin{figure}[bth!]
\includegraphics[scale=0.28]{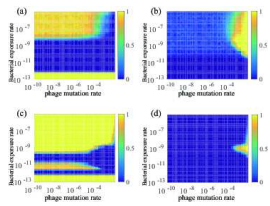}
\caption{\label{extinction2} (Color Online) The extinction probability of
\textbf{(a)} phages and \textbf{(b)} bacteria.
Here the probability of acquiring new
spacers is $\gamma = 0.0005$ and the rate of
deleting one spacer is $d = 0.0001\ \mathrm{min}^{-1}$.
The extinction probability of
\textbf {(c)}
phages
and \textbf {(d)}
bacteria.
Here $\gamma=0.005$ and $d = 10^{-5}\ \mathrm{min}^{-1}$.
Other parameters are the same as those in Fig.~\ref{extinction1}.}
\end{figure}

From the above explanation of the four regions in Fig.~\ref{extinction1}, we have the conditions for which this interesting non-classical phase diagram of phage extinction exists. First, bacteria must possess the CRISPR adaptive immune system: if bacteria do not have CRISPR, bacteria and phages can only coexist when $\beta$ is small, $\beta \approx 10^{-12}\ \mathrm{mL}\cdot\mathrm{min}^{-1}$, and region (2) and region (4) will not exist. Second, bacteria must have some rate to lose the acquired immunity. If bacteria can accumulate an unlimited number of spacers, phages will eventually go extinct if the length of the proto-spacers is finite and region (3) will not exist. Third, the rate of losing the adaptive immunity must be small. In region (2) and the left boundary of region (4), phages cannot grow back because the rate of losing spacers is small. If the rate of losing spacers is large, region (2) will disappear and the left boundary of region (4) will move towards higher $\beta$ values, as shown in Fig.~\ref{extinction2}(a) and (b). Conversely this phase diagram is not sensitive to the probability of acquiring new spacers. Increasing $\gamma$ only changes the pattern of the extinction probability in high $\mu$ regions, making it more difficult for phages to escape from CRISPR recognition, as shown in Fig.~\ref{extinction2}(c) and (d). From the above results, we predict when the deletion rate of spacer and the mutation rate of phages is small, decreasing the adsorption rate of phages can make phages extinct. However, further decreasing the adsorption rate can allow phages to reemerge.

\begin{figure}[hbt!]
\includegraphics[scale=0.36]{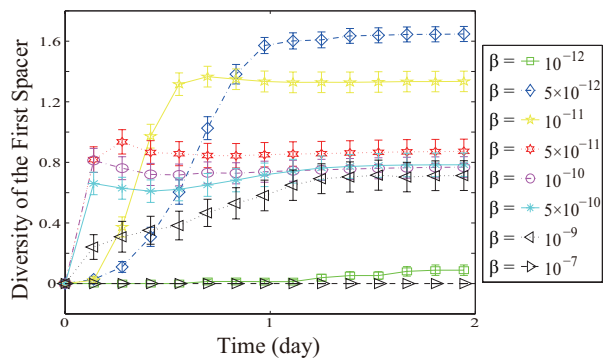}
\caption{\label{SpacerDiversity} (Color Online) The diversity of the first spacer, defined as $D = - \sum_{k} p_{s_0}(k) \log p_{s_0}(k) $, where $p_{s_0}(k)$ is the observed probability to have sequence $k$ at the first position, $s_0$, in the spacer array in CRISPR, for different values of bacterial
exposure rate $\beta$ when the mutation rate is $\mu=10^{-8}$. The other parameters are the same as those in Fig.~\ref{extinction1}.}
\end{figure}

CRISPR changes the bacterial population structure.
In Fig.~\ref{SpacerDiversity}, the Shannon entropy of the first spacer is used as a measure of the diversity.
In Fig.~\ref{SpacerDiversity}, the diversity of spacers rises slowly when $\beta$ is small, region (1) in Fig.~\ref{extinction1}(c). This is because the selection pressure on bacteria is small, and CRISPR does not provide bacteria much advantage. As $\beta$ increases, the diversity of spacers increases faster because the density of phages is larger and the value of $\beta$ is higher, making the adsorption of phages into bacteria more rapid. But the steady-state value of the diversity decreases, implying the distribution of spacers becomes more biased. If the selection pressure on bacteria is larger, the bacteria that acquire spacers will dominate the population in a shorter time. When the bacteria that have spacers dominate the population, phages are repressed, and the density of phages stays at a low level. The process of acquiring spacers becomes slower, leading to a smaller steady value of spacer diversity.

\begin{figure}[htb!]
\includegraphics[scale=0.27]{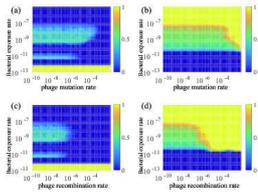}
\caption{\label{RecombinationMutation} (Color Online) The extinction probability
of \textbf{(a)} phages and \textbf{(b)} bacteria
when phages undergo only point mutation. The extinction probability of \textbf{(c)} phages
and \textbf{(d)} bacteria when phages undergo only recombination. Initially, there are two strains of phages. Here the probability of acquiring new spacers is $\gamma = 0.01$. The other parameters are the same as those in Fig.~\ref{extinction1}.}
\end{figure}

Phages can have rapid recombination \cite{Andersson2008}. Recombination is compared to point mutation of phages in Fig.~\ref{RecombinationMutation}.
Here there are two strains of phages initially, and so
acquisition of two spacers is required for bacterial immunity.
The limiting extinction probability in this case is $1-(1-\gamma^2)^{N^B_0} \approx 0.40$ in Fig.~\ref{RecombinationMutation}. Additionally, at very large $\beta$, bacteria with a finite number of spacers, $n_s$, eventually go extinct when the spacer array by chance is entirely occupied by proto-spacers from only one strain of phage. Finally, the extinction probability of phages when phages have only recombination is lower than that when phages have only point mutation, because recombination can change several proto-spacers at once.

\section{Discussion}

The cost of adding novel spacers is undetectable in some experiments \cite{Westra_Current_Biology,ValeLittle2010}. In our model, we set the cost of adding new spacers to a small value, consistent with the experimental data. We also found that the results are persistent with changes to the cost of adding novel spacers in our model. When we set the cost of adding new spacers to zero and $0.5$, the results, which are shown in Fig.~\ref{fig:cost_0} and Fig.~\ref{fig:cost_0.5}, are almost identical with those when the cost of adding new spacers is $0.1$. 

Here we showed that bacteria can coexist with one phage strain because of the balance
between acquisition and deletion of spacers. But this balance cannot always be achieved, and in some parameter regimes either the phages or bacteria go extinct.
For example, the study of \cite{Paez2013} showed coexistence of phage and
bacteria, wereas the study of \cite{Houte2016} showed elimination of phage
by bacteria for a sufficiently diverse bacteria population of CRISPR spacers.
  When the bacterial exposure rate varies, the coexistence of bacteria and phages shows an interesting pattern of reentrant phases.
Thus, a testable prediction of our model is that
when the bacterial exposure rate is low, phages go to extinction; increasing the bacterial exposure rate makes the phage population emerge in the system, but increasing the bacterial exposure rate still further can result in phages extinction. Phages can further reemerge if the bacterial exposure rate is increased more. Finally, phages go extinct when the bacterial exposure rate excesses a critical threshold. The whole process is depicted in Fig.~\ref{extinction1}(c). 
The bacterial exposure rate may 
change due to the variation in phage tail and host receptor affinities, or because of the change in temperature and ion densities, and changes to this exposure rate strongly influence the 
balance between acquisition and deletion of spacers.

When there is greater than a single phage sequence, for example, when the phage mutation rate increases, the coexistence of bacteria and phages is stabilized due to less ability of the bacteria to recognize the diverse phage strains. Here our approach has mimicked
controlled environments such as laboratory and factory strains
rather than natural environment strains such as those arising in the ocean \cite{Flores2013,Haerter2014}. In natural
environments, the diversity of bacteria and phages is likely rather large. In our current model, there is initially a single bacteria strain and one or two phages strains initially. As time elapses, the diversity of bacteria increases, but the diversity of phages remains low because the phage mutation rate is relatively small.

The boundaries of the phases that arose from the stochastic co-evolution
process were explained by a mean-field analysis. In this way,
we theoretically estimated the threshold of the bacterial exposure rate, $\beta$, at which the bacteria and phages can coexist and gained the insight
into why bacteria and phages coexist. The phase diagrams showed here
are the results at steady state, when
average the total densities of bacteria and phages remain unchanging
with time, with the density of each species fluctuating around the
average values.

When multiple species of phage were present, recombination allowed
the phage to more easily escape extinction by the CRISPR immune
system. The phase diagram was shifted such that lower rates of 
recombination were as effective at immune evasion as were higher
rates of mutation.  These results support the interpretation of
long-term bacterium-phage coevolution experiments,
in which recombination among multiple phage strains enable phage
persistance against the bacterial CRISPR system \cite{Paez2015}.

Other properties of the phage-bacteria coevolving system also
affect the phase boundaries.  For example, when the spacer
diversity is sufficient, the phages can be driven to extinction.
The boundary for extinction depends on the number of spaces
in the CRISPR system, as seen for example, by comparing the present
results to previous results for a larger CRISPR  array
\cite{Pu2013}.  The ability of CRISPR to drive phages to extinction
has been seen experimentally \cite{Houte2016}.

We note that high rates of bacterial exposure lead to phage persistance and
bacterial extinction.  High rates of exposure may result from effect
contact between the phage and bacteria.  High rates may also result from
migration of naive bacteria to regions of high phage concentration and
diversity.  From the present results, we see that
CRISPR will become a less effective protection mechanism at high exposure or
migration rates.  This result has been obeserved experimentally, where
high bacterial migration rates induced a shift from CRISPR-mediated
protection to a surface modification-mediated defense by bacteria
\cite{Chabas2016}.  The more specific CRISPR mechanism is effective
when bacteria have enough time to incorporate the proto-spacers
providing protection \cite{Pu2013,Westra_Current_Biology}.

In summary, we predict an interesting phase diagram of phage extinction probability. When the deletion rate of spacers in CRISPR is small, phages go extinct when the value of $\beta$ is low, but phages can coexist with bacteria when $\beta$ is even lower.
CRISPR changes the evolution of bacteria and phages, accelerating the co-evolution of bacteria and phages. Finally, recombination is a more efficient mechanism for phages to escape the recognition of CRISPR than is point mutation when there are multiple proto-spacers in the phage.
Future work may consider biotechnology applications,
genome editing approaches, population-level bacterial control,
or effects of recombination in the microbiome.

\section*{Authors' contribution}
P.H. wrote the codes, collected and analyzed the data, drafted the manuscript. Both P.H. and M.W.D developed and analyzed the model. M.W.D helped to draft the manuscript. All authors gave final approval for publication.

\section*{Competing interest}
We declare we have no competing interests.

\section*{Funding}
We received no funding for this study.

\section*{Acknowledgment}
We thank Dr Jeong-Man Park, the Catholic University of Korea, for helpful
discussions about the method of this paper.

\section*{Appendix}

\setcounter{figure}{0}
\setcounter{equation}{0}
\renewcommand{\thefigure}{A\arabic{figure}}
\renewcommand{\theequation}{A\arabic{equation}}
\renewcommand{\thetable}{A\arabic{table}}
\newcommand{\pvec}[1]{\vec{#1}\mkern2mu\vphantom{#1}}

\subsection{Table of Parameters}
The parameters used in the main text are listed in table.~\ref{TableofParameters}.

\begin{table}[!ht]
\centering
\caption{\bf Table of Parameters}\label{TableofParameters}
\begin{tabularx} {\textwidth} {c | X | c | c }
\hline
Parameter & Meaning & Value & References\\
\hline
$c_0$ & Bacterial growth rate & 0.005 $\mathrm{min}^{-1}$ & \cite{Paez2013}\\ \hline
$c$  & Cost of each spacer & 0.1 & \cite{Jiang2013}\\ \hline
$\beta$ & Bacterial exposure rate & $10^{-13}$--$10^{-5}$ mL $ \cdot \mathrm{min}^{-1}$ & \cite{Moldovan2007}\\ \hline
$\gamma$ & Probability of acquiring new spacers & 0.0005 & \cite{Childs2011}, \cite{Weinberger2012}\\ \hline
$d$ & Rate of deleting one spacer  & $10^{-5} \mathrm{min}^{-1}$ & \cite{Jiang2013}\\ \hline
$\tau$ & Latent time & 40 min & \cite{Ellis1939}\\ \hline
$b$ & Phage burst size & 100 & \cite{Ellis1939}\\ \hline
$\delta$ & Phage decay rate & 0.001 $\mathrm{min}^{-1}$ & \cite{Shttle1992} \\ \hline
$\mu$ & Mutation rate & $10^{-10}$--$10^{-2}$ per base per replication & \cite{Rafael2010}\\ \hline
$\nu$ & Recombination rate & $10^{-10}$--$10^{-2}$ per base per replication & \cite{Benbow1974}\\ \hline
$n_p$ & Number of proto-spacers in phages & 30 & \cite{Paez2013}\\ \hline
$n_s$ & Maximum number of spacers in CRISPR & 6 & \cite{Paez2013}\\ \hline
$L$ & Length of each spacer and proto-spacer & 10 bp& \cite{Semenova2011}\\ \hline
$x^B_{_\mathrm{M}}$ & Maximum bacterial density & $10^9 \mathrm{mL}^{-1}$ & \cite{Paez2013}\\ \hline
$V$ & Volume of the system  & $10^{-3} \mathrm{mL}^{-1}$ \\
\hline
\end{tabularx}
\end{table}
The values of $c_0$ and $x^B_{_\mathrm{M}}$ are estimated from \cite{Paez2013}.
The cost of spacers is low \cite{Jiang2013}; here we choose $c=0.1$.
The values of $b$ and $\tau$ are estimated from \cite{Ellis1939}.
The value of $\beta$ is estimated from \cite{Moldovan2007}.
The value of $\gamma$ is estimated from \cite{Childs2011} and \cite{Weinberger2012}.
The value of $d$ is estimated from \cite{Jiang2013}.
The value of $\delta$ is estimated from \cite{Shttle1992}.
The value of $\mu$ is estimated from \cite{Rafael2010}.
The value of $\nu$ is estimated from \cite{Benbow1974} and is the same order as the value of $\mu$.
The interference between proto-spacers and CRISPR spacers is governed by the PAM and the seed regions \cite{Semenova2011}. The length of the PAM is about 3 bp and the length of the seed region is 7 bp \cite{Semenova2011}, so we set the length of spacers and proto-spacers to 10 pb.
In the experiment to which we compare \cite{Paez2013}, the average number of spacers in CRISPR is small, on average 0.8 spacers per bacteria, so we set the maximum number of spacers to 6. The average number of spacers in our simulation is shown in Fig.~\ref{spacer_num_1}.
There are 27 spacers that account for between 82\% and 99\% of all spacers sampled on any individual day in the experiment to which we compare \cite{Paez2013}. Here we set $n_p$ to 30. We also tried $n_p=1$. The results are qualitatively the same, as shown in Fig.~\ref{supp_change_protospacer}.
The volume $V$ is set to mimic the typical volume of a droplet.

\subsection{Master Equation}

The master equation of the stochastic process described in the main text is

\begin{eqnarray}\label{MasterEq}
& &\frac{d P(\{N^B_{\vec{s}},N^I_{\vec{p}_1},N^P_{\vec{p}_2}\})}{dt}
= \sum_{\vec{s}} \frac{c_0}{1 + c \cdot k_{\vec{s}} } (N^B_{\vec{s}} -1 ) \Big\{ 1 - \Big[ (\sum_{\pvec{s}'} N^B_{\pvec{s}'}) -1 + \sum_{\vec{p}} N^I_{\vec{p}} \Big]/N^B_{_{\mathrm{M}}} \Big\}  \nonumber \\
& & P(N^B_{\vec{s}}-1)
- \sum_{\vec{s}} \frac{c_0}{1 + c \cdot k_{\vec{s}} } N^B_{\vec{s}}  \Big[ 1 - (\sum_{\pvec{s}'} N^B_{\pvec{s}'} + \sum_{\vec{p}} N^I_{\vec{p}} )/N^B_{_{\mathrm{M}}} \Big] P(N^B_{\vec{s}}) \nonumber \\
&  & + \frac{(1-\gamma) \beta}{V} \sum_{\vec{s}} \sum_{\vec{p}_2} (N^P_{\vec{p}_2}+1) (N^B_{\vec{s}}+1) \theta_1(\vec{p}_2,\vec{s}) P(N^B_{\vec{s}}+1,N^I_{\vec{p}_2}-1,N^P_{\vec{p}_2}+1) \nonumber \\
& & + \frac{(1-\gamma) \beta}{V} \sum_{\vec{s}} \sum_{\vec{p}_2} (N^P_{\vec{p}_2}+1) N^B_{\vec{s}} \left[1-\theta_1(\vec{p}_2,\vec{s})\right] P(N^B_{\vec{s}},N^P_{\vec{p}_2}+1) \nonumber \\
& & - \frac{\beta}{V} \sum_{\vec{s}} \sum_{\vec{p}_2} N^P_{\vec{p}_2} N^B_{\vec{s}} P(N^B_{\vec{s}},N^P_{\vec{p}_2}) + \frac{\gamma\beta}{V \cdot n_p} \nonumber \\
& & \sum_{\vec{s}} \sum_{\pvec{s}'} \sum_{\vec{p}_2} \sum_{i=1}^{n_p} (N^P_{\vec{p}_2}+1) (N^B_{\vec{s}'}+1) \theta_2(p_{2_i},\pvec{s}',\vec{s})P(N^B_{\pvec{s}'}+1,N^B_{\vec{s}}-1,N^P_{\vec{p}_2}+1)
\nonumber \\
& & + d \sum_{\vec{s}}  \sum_{\pvec{s}'} (N^B_{\pvec{s}'}+1) \sum_{i=1}^{k_{\pvec{s}'}}
\theta_3(s_i,\pvec{s}',\vec{s}) P(N^B_{\pvec{s}'}+1,N^B_{\vec{s}}-1)
- d \sum_{\vec{s}} k_{\vec{s}} N^B_{\vec{s}} P(N^B_{\vec{s}}) \nonumber \\
& & + \delta \sum_{\vec{p}_2}  (N^P_{\vec{p}_2}+1) P(N^P_{\vec{p}_2}+1) -  \delta \sum_{\vec{p}_2}  N^P_{\vec{p}_2} P(N^P_{\vec{p}_2})\nonumber \\
& & +  \frac{1-\nu}{\tau} \sum_{\vec{p}_1} \sum_{\pvec{p}_1',\ldots,\pvec{p}_b'}  (N^I_{\vec{p}_1} + 1)  \prod^{b}_{i=1} \left( \mu^{h(\vec{p}_1,\pvec{p}_i')}  (1-\mu)^{L\cdot n_p-h(\vec{p}_1,\pvec{p}_i')}\right)\nonumber\\
& &\cdot P(N^I_{\vec{p}_1} + 1,N^P_{\pvec{p}_1'}-1,\ldots,N^P_{\pvec{p}_b'}-1) + \frac{\nu}{\tau[(\sum_{\vec{p}}N^P_{\vec{p}})-b]^b}  \nonumber \\
& & \sum_{\vec{p}_1} \sum_{\pvec{p}_1^*,\ldots,\pvec{p}_b^*} \sum_{\pvec{p}_1',\ldots,\pvec{p}_b'} \sum_{\pvec{p}_1'',\cdots,\pvec{p}_b''}(N^I_{\vec{p}_1}+1) \prod_{i=1}^{b}\Big[\mu^{h(\vec{p}_1,\pvec{p}_i')} (1-\mu)^{L\cdot n_p - h(\vec{p}_1,\pvec{p}_i')}   \nonumber \\
& & \cdot (N^P_{\pvec{p}_i^*}- \sum_{j=1}^b{\Delta_{\pvec{p}_i^*,\pvec{p}_j''}}) \sum_{\vec{r}} \prod_{k=1}^{L\cdot n_p}  p_c^{{\vec{r}_k}} (1-p_c)^{1-{\vec{r}_k}} \theta_4(\pvec{p}_i',\pvec{p}_i^*,\pvec{p}_i'',\vec{r})\Big] \nonumber \\
& & \cdot P(N^I_{\vec{p}_1}+1,N^P_{\pvec{p}_1''}-1,\ldots,N^P_{\pvec{p}_b''}-1) - \frac{1}{\tau} \sum_{\vec{p}_1} N^I_{\vec{p}_1} P(N^I_{\vec{p}_1})
\end{eqnarray}
where $N^B_{\vec{s}}$ is the population of the bacteria with spacer array $\vec{s}$, $N^I_{\vec{p}_1}$ is the population of infected bacteria invaded by phages with proto-spacer array $\vec{p}_1$, $N^P_{\vec{p}_2}$ is the population of phages with proto-spacer array $\vec{p}_2$, and $N^B_{_{\mathrm{M}}}$ is the maximum population of bacteria.
In Eq.~\ref{MasterEq}, $\theta_1(\vec{p},\vec{s}) = 0$ when $\vec{s}$ recognizes ${\vec{p}}$ and 1 otherwise. The $\theta_2(p_i,\pvec{s}',\vec{s})$ $=$ 1 when $\{p_i,s'_1,\ldots,s'_{n_s -1}\}$ = $\vec{s}$ and 0 otherwise. The $\theta_3(s'_i,\pvec{s}',\vec{s})$ $=$ 1 when $\{s'_1,\ldots,s'_{i-1},s'_{i+1},\ldots,s'_{n_s},0\}$ = $\vec{s}$ and 0 otherwise. The hamming distance between $\vec{p}_1$ and $\pvec{p}_i'$ is $h(\vec{p}_1,\pvec{p}_i')$.
The $\vec{r}$ is a bit string, which denotes the recombination pattern. Each bit in $\vec{r}$ is either 0 or 1. If $\vec{r}_k=1$, it means there is a crossover at position $k$. If $\vec{r}_k=0$, it means there is no crossover at position $k$.
The $\theta_4(\pvec{p}_i',\pvec{p}_i^*,\pvec{p}_i'',\vec{r})=1$ if $\pvec{p}_i''$ can be generated by the recombination pattern $\vec{r}$ from $\pvec{p}_i'$ and $\pvec{p}_i^*$ and 0 otherwise.
$\Delta_{\pvec{p}_i,\pvec{p}_j'}=1$ if $\pvec{p}_i = \pvec{p}_j'$ and 0 otherwise. In $P(N^I_{\vec{p}_1} + 1,N^P_{\pvec{p}_1'}-1,\ldots,N^P_{\pvec{p}_b'}-1)$, if $\pvec{p}_i' = \pvec{p}_j'$, it means
$N^P_{\pvec{p}_i'}-2$. In general, $P(N^I_{\vec{p}_1} + 1,N^P_{\pvec{p}_1'}-1,\ldots,N^P_{\pvec{p}_b'}-1)$ is
short hand for $ P(N^I_{\vec{p}_1} + 1,\{N^P_{\pvec{p}'_k}-\sum_{j=1}^b \Delta_{\pvec{p}'_k,\pvec{p}_j'}\})$.

We can show that
\begin{equation*}
\sum_{\pvec{p}_1',\ldots,\pvec{p}_b'} \prod^{b}_{i=1} \left( \mu^{h(\vec{p}_1,\pvec{p}_i')}  (1-\mu)^{L\cdot n_p-h(\vec{p}_1,\pvec{p}_i')}\right) = 1
\end{equation*}
and
\begin{eqnarray*}
& & \sum_{\pvec{p}_i''} \sum_{\vec{r}} \prod_{k=1}^{L\cdot n_p}  p_c^{{\vec{r}_k}} (1-p_c)^{1-{\vec{r}_k}} \theta_4(\pvec{p}_i',\pvec{p}_i^*,\pvec{p}_i'',\vec{r}) \nonumber \\
& = & \sum_{\vec{r}} \prod_{k=1}^{L\cdot n_p}  p_c^{{\vec{r}_k}} (1-p_c)^{1-{\vec{r}_k}} \sum_{\pvec{p}_i''} \theta_4(\pvec{p}_i',\pvec{p}_i^*,\pvec{p}_i'',\vec{r}) \nonumber \\
& = & \sum_{\vec{r}} \prod_{k=1}^{L\cdot n_p}  p_c^{{\vec{r}_k}} (1-p_c)^{1-{\vec{r}_k}}\nonumber \\
& = & 1
\end{eqnarray*}

Therefore,
\begin{eqnarray*}
& & \sum_{\pvec{p}_1^*,\ldots,\pvec{p}_b^*} \sum_{\pvec{p}_1',\ldots,\pvec{p}_b'} \sum_{\pvec{p}_1'',\ldots,\pvec{p}_b''}\prod_{i=1}^{b}\Big[\mu^{h(\vec{p}_1,\pvec{p}_i')} (1-\mu)^{L\cdot n_p - h(\vec{p}_1,\pvec{p}_i')} N^P_{\pvec{p}_i^*} \sum_{\vec{r}} \prod_{k=1}^{L\cdot n_p} \\
& & p_c^{{\vec{r}_k}} (1-p_c)^{1-{\vec{r}_k}} \theta_4(\pvec{p}_i',\pvec{p}_i^*,\pvec{p}_i'',\vec{r}) \Big] = (\sum_{\vec{p}}N^P_{\vec{p}})^b
\end{eqnarray*}
and
\begin{eqnarray*}
& &- \frac{1-\nu}{\tau} \sum_{\vec{p}_1} \sum_{\pvec{p}_1',\ldots,\pvec{p}_b'} N^I_{\vec{p}_1} \prod^{b}_{i=1} \left( \mu^{h(\vec{p}_1,\pvec{p}_i')}  (1-\mu)^{L\cdot n_p-h(\vec{p}_1,\pvec{p}_i')}\right) P(N^I_{\vec{p}_1}) \\
& & -  \frac{\nu}{\tau (\sum_{\vec{p}}N^P_{\vec{p}})^b} \sum_{\vec{p}_1} \sum_{\pvec{p}_1^*,\ldots,\pvec{p}_b^*} \sum_{\pvec{p}_1',\ldots,\pvec{p}_b'} \sum_{\pvec{p}_1'',\ldots,\pvec{p}_b''} N^I_{\vec{p}_1}
\prod_{i=1}^{b}\Big[\mu^{h(\vec{p}_1,\pvec{p}_i')} (1-\mu)^{L\cdot n_p - h(\vec{p}_1,\pvec{p}_i')} N^P_{\pvec{p}_i^*} \\
& & \sum_{\vec{r}} \prod_{k=1}^{L\cdot n_p}  p_c^{{\vec{r}_k}} (1-p_c)^{1-{\vec{r}_k}} \theta_4(\pvec{p}_i',\pvec{p}_i^*,\pvec{p}_i'',\vec{r}) \Big] P(N^I_{\vec{p}_1})
=  - \frac{1}{\tau} \sum_{\vec{p}_1} N^I_{\vec{p}_1} P(N^I_{\vec{p}_1})
\end{eqnarray*}
This is why we get the last term in Eq.~\ref{MasterEq}.

\subsection{Mean Field Equations}
The corresponding mean field equations for the densities of bacteria and phages, shown for illustrative purpose and not used in the simulations
reported in the main text, are

\begin{subequations}\label{MeanFieldEqs}
\begin{eqnarray}
& & \frac{d x^B_{\vec{s}}}{dt} = \frac{c_0}{1+ c \cdot k_{\vec{s}} } x^B_{\vec{s}}  \big [ 1 - (\sum_{\pvec{s}'} x^B_{\pvec{s}'} +
\sum_{\vec{p}} x^I_{\vec{p}})/x^B_{_{\mathrm{M}}} \big]
- (1 - \gamma) \beta x^B_{\vec{s}} \sum_{\vec{p}} x^P_{\vec{p}}
\theta_1 (\vec{p},\vec{s}) - {\gamma \beta}x^B_{\vec{s}}  \nonumber \\
& & \sum_{\vec{p}}  x^P_{\vec{p}}
+ \frac{\gamma \beta}{n_p}\sum_{\pvec{s}'} \sum_{\vec{p}} \sum_{i=1}^{n_p} x^B_{\pvec{s}'} x^P_{\vec{p}} \theta_2(p_i,\pvec{s}',\vec{s})
- d \cdot k_{\vec{s}} x^B_{\vec{s}} + d \sum_{\pvec{s}'} x^B_{\pvec{s}'} \sum_{i=1}^{k_{\pvec{s}'}}  \theta_3 (s'_i,\pvec{s}',\vec{s}), \label{eq2.1}
\end{eqnarray}

\begin{eqnarray}
\frac{d x^I_{\vec{p}}}{dt} & = &  (1 - \gamma) \beta \sum_{\vec{s}} x^B_{\vec{s}} x^P_{\vec{p}}  \theta_1(\vec{p},\vec{s})
 -  \frac{x^I_{\vec{p}}}{\tau}, \label{eq2.2}
\end{eqnarray}

\begin{eqnarray}
& & \frac{d x^P_{\vec{p}}}{dt} = \frac{b(1-\nu)}{\tau} \sum_{\pvec{p}'}x^I_{\pvec{p}'}
\mu^{h(\vec{p},\pvec{p}')}(1-\mu)^{L\cdot n_p-h(\vec{p},\pvec{p}')}
+ \frac{b \nu}{\tau\sum_{\pvec{p}'}x^P_{\pvec{p}'}} \sum_{\vec{p_1}}\sum_{\vec{p_2}}\sum_{\vec{p_3}} x^I_{\vec{p_1}}
\mu^{h(\vec{p}_1,\vec{p_2})} \nonumber \\ & &(1-\mu)^{L\cdot n_p-h(\vec{p}_1,\vec{p_2})}
\cdot x^P_{\vec{p_3}} \sum_{\vec{r}} \prod_{k=1}^{L\cdot n_p} p_c^{\vec{r}_k} (1-p_c)^{1-\vec{r}_k}
\theta_4(\vec{p_2},\vec{p_3},\vec{p},\vec{r}) \nonumber \\
& & -  \delta \cdot x^P_{\vec{p}}
- \beta x^P_{\vec{p}} \sum_{\vec{s}}  x^B_{\vec{s}}, \label{eq2.3}
\end{eqnarray}
\end{subequations}
where $x^B_{\vec{s}}$ is the density of bacteria with spacer array $\vec{s}$, $x^I_{\vec{p}}$ is the density of infected bacteria invaded by phages with proto-spacer array $\vec{p}$, and $x^P_{\vec{p}}$ is the density of phages with proto-spacer array $\vec{p}$. The functions of $\theta_1$, $\theta_2$, $\theta_3$ and $\theta_4$ are the same as those in Eq.~\ref{MasterEq}.

\subsection{Varying the Cost of Adding New Spacers}

The phase diagrams of the extinction probability of phages and bacteria do not change when the cost of adding novel spacers varies. 

\begin{figure}[hbt!]
\centering
\includegraphics[width=0.65\textwidth]{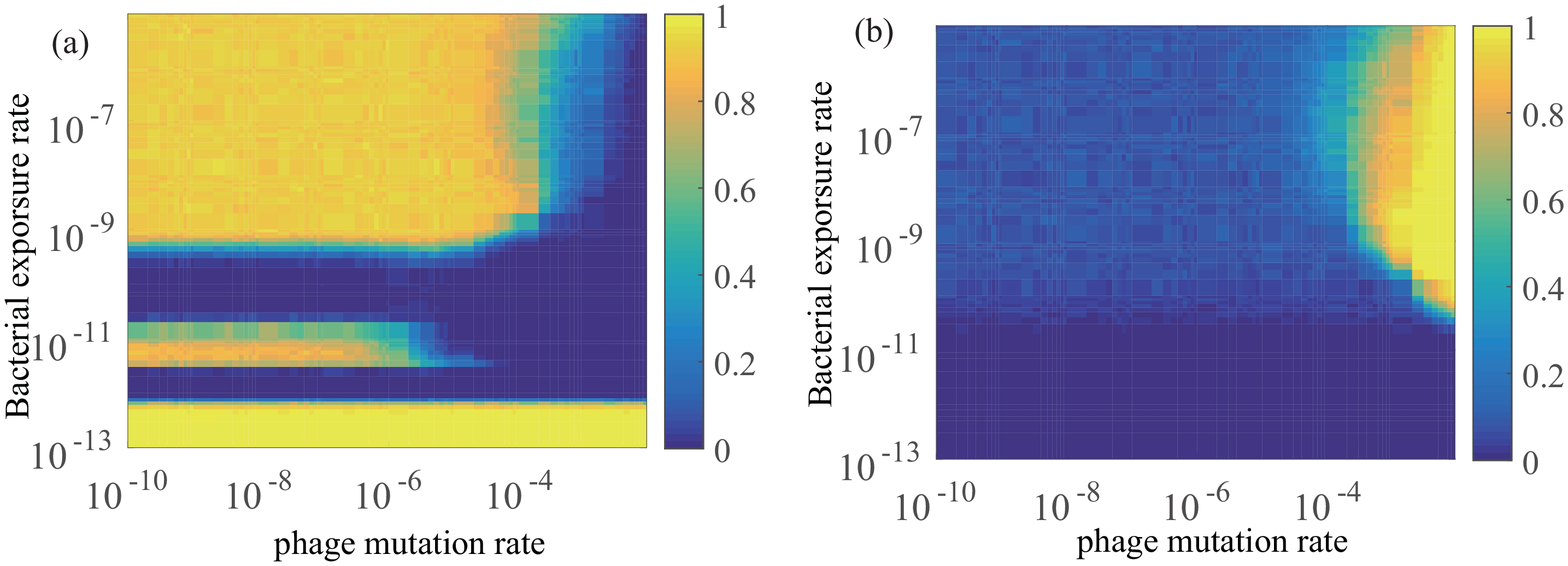}
\caption{\label{fig:cost_0}  The extinction probability of
\textbf{(a)} phages and \textbf{(b)} bacteria.
Here the cost of adding new spacers is zero.
Other parameters are the same as those in Fig.~\ref{extinction1} in the main text.}
\end{figure}

\begin{figure}[hbt!]
\centering
\includegraphics[width=0.65\textwidth]{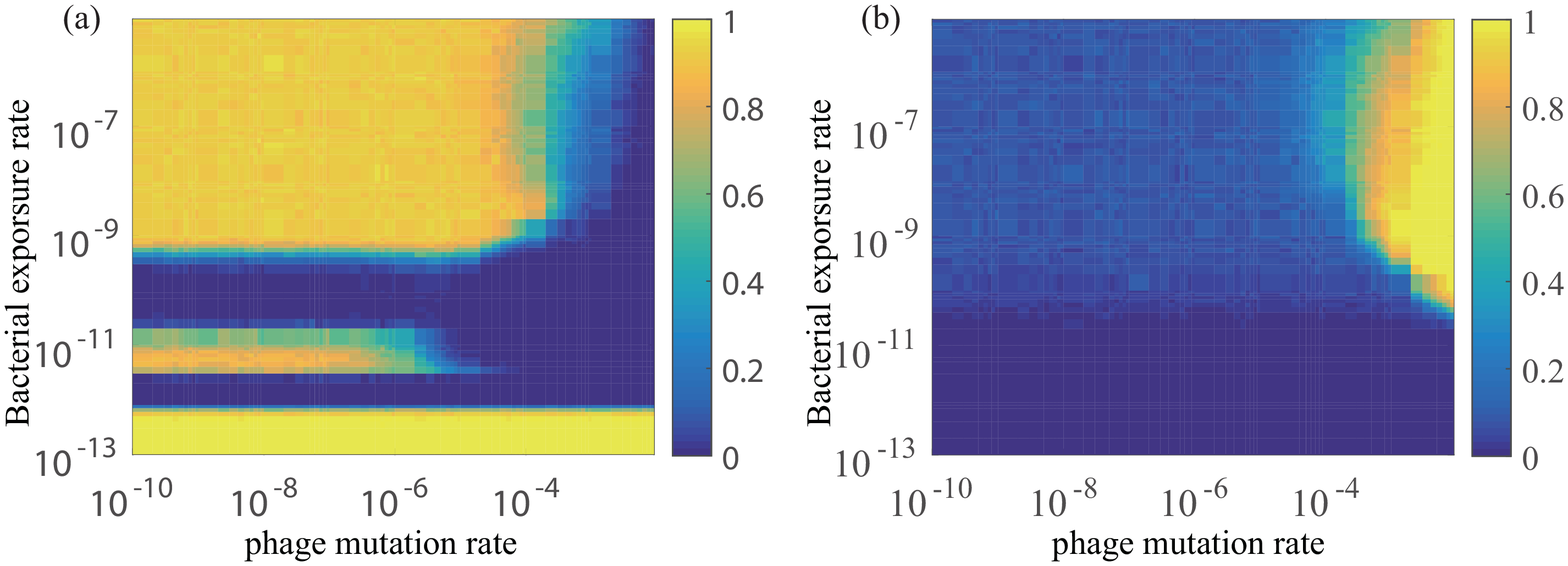}
\caption{\label{fig:cost_0.5}  The extinction probability of
\textbf{(a)} phages and \textbf{(b)} bacteria.
Here the cost of adding new spacers is $0.5$.
Other parameters are the same as those in Fig.~\ref{extinction1} in the main text.}
\end{figure}

\subsection{Number of Spacers}

The average number of spacers in our simulation does not reach $n_s$ in most of the parameter regime. In the range $\beta \in [10^{-12}, 10^{-8}]$ and $\nu \in [10^{-8},10^{-6}]$, the average number of spacers is 0--2, which is in agreement with the experiment data in \cite{Paez2013}.

\begin{figure}[htb!]
\includegraphics[width=0.65\textwidth]{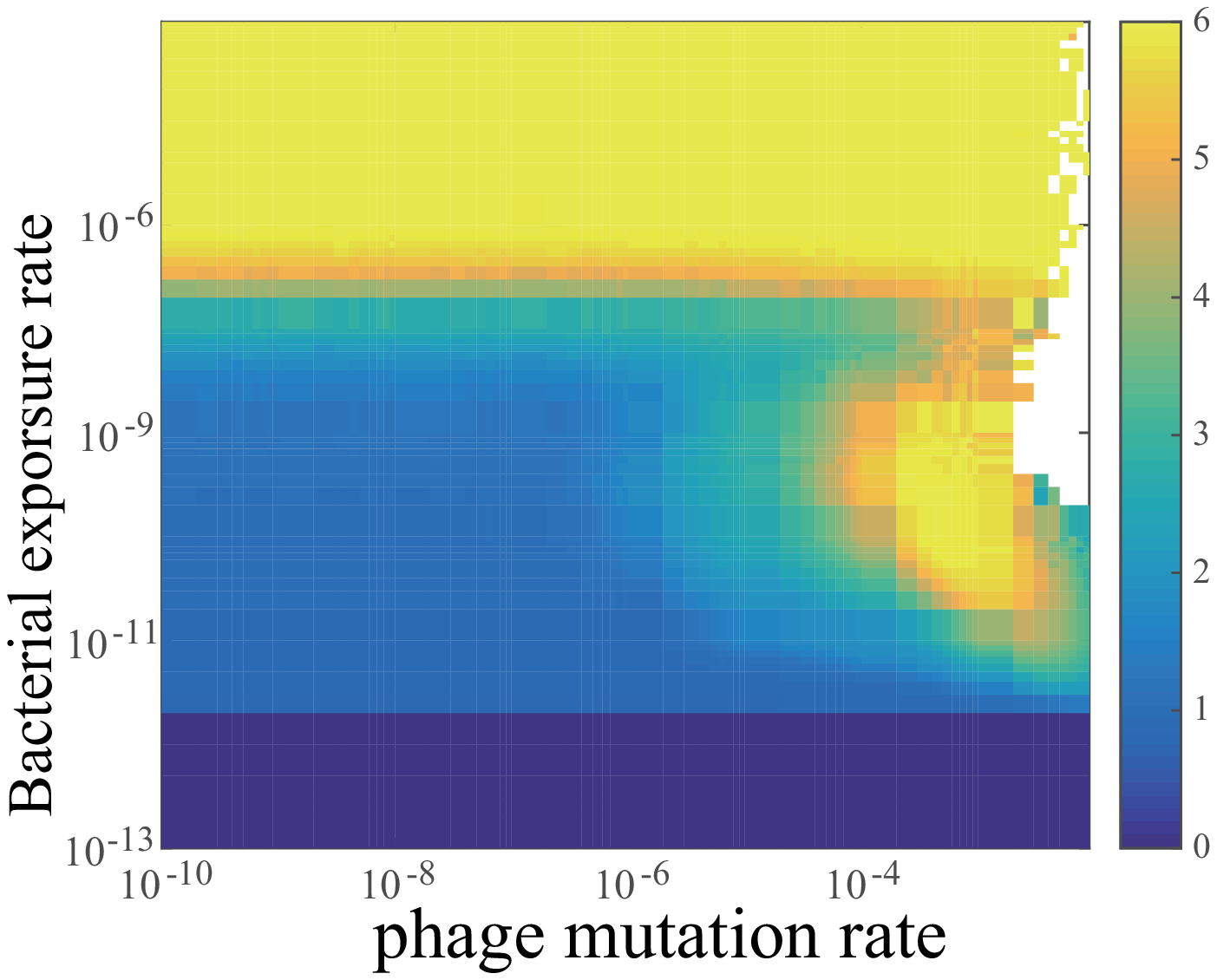}
\caption{\label{spacer_num_1} The average number of spacers in CRISPR. The parameters are the same as those in Fig.~1 in the main text. Blank means no data available.}
\end{figure}

\begin{figure}[htb!]
\includegraphics[width=0.65\textwidth]{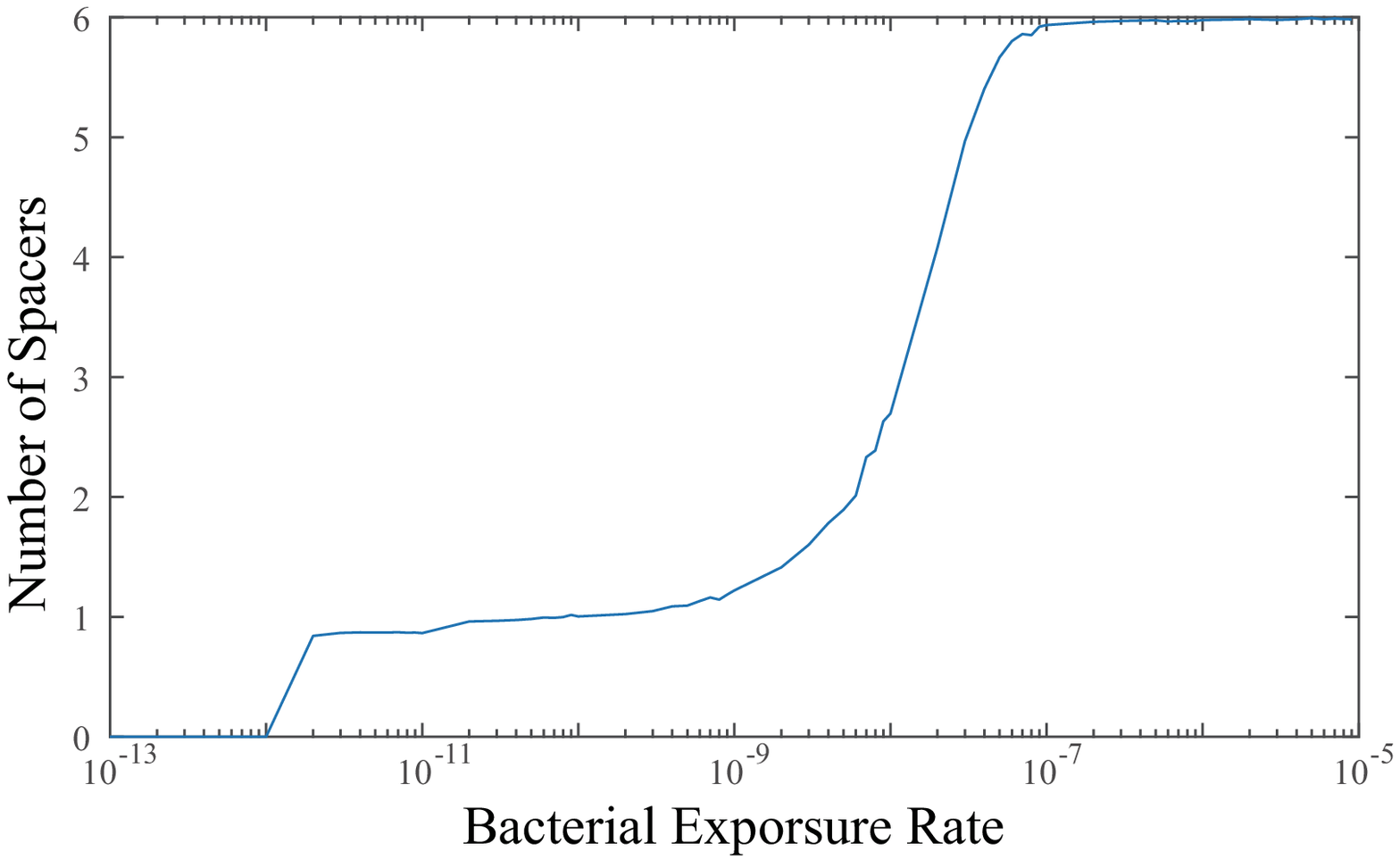}
\caption{\label{spacer_num_2} The average number of spacers in CRISPR when $\mu = 10^{-8}$ in Fig.~\ref{spacer_num_1}. The parameters are the same as those in Fig.~1 in the main text.}
\end{figure}

\subsection{Number of Proto-spacers}

When $n_p=1$, the pattern of the extinction probability of phages is qualitatively the same as Fig.~1 in the main text.

\begin{figure}[htb!]
\includegraphics[width=0.85\textwidth]{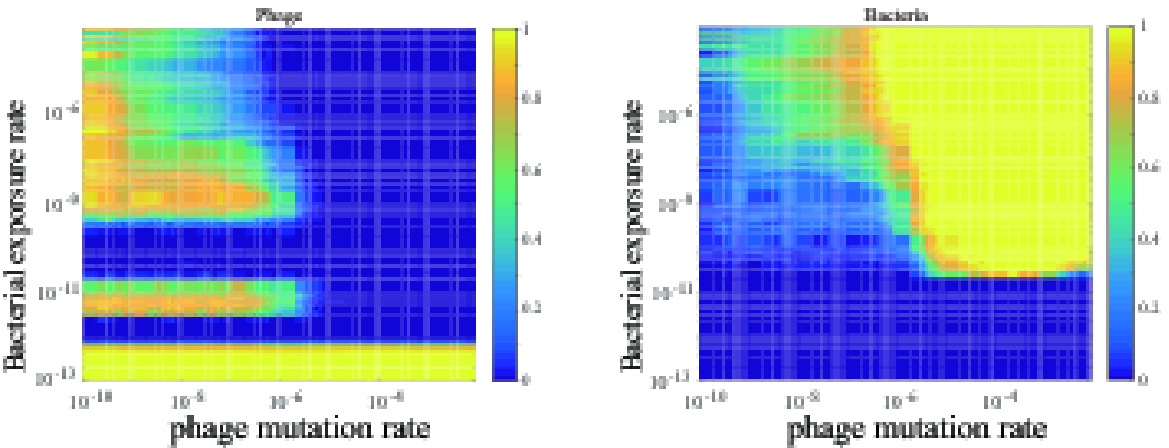}
\caption{\label{supp_change_protospacer} The extinction probability of bacteria
and phages when $n_p=1$. Other parameters are the same as those in Fig.~1 in the main text.}
\end{figure}


Author contributions: MWD conceived of the study and wrote the manuscript. PH carried out the research and wrote the manuscript.

\bibliographystyle{vancouver}
\bibliography{crispr}

\end{document}